\documentclass[conference, 10pt]{IEEEtran}

\usepackage{times}
\usepackage[latin1]{inputenc}
\usepackage[T1]{fontenc}
\usepackage[english]{babel}
\usepackage{amsmath, amssymb, graphicx, array, tabularx,booktabs}
\usepackage{epsf,epsfig}
\usepackage{setspace}
\usepackage{subfigure}
\usepackage{algorithmic}
\usepackage{algorithm}
\usepackage{multirow}
\usepackage{url}
\usepackage{color}
\usepackage[nolist,printonlyused]{acronym}
\usepackage{comment}
\usepackage{authblk}
\usepackage{tabularx}
\usepackage{siunitx}
\usepackage[font=small,skip=0pt]{caption}
\def\r1#1{\textcolor{black}{#1}}
\def\q2#1{\textcolor{black}{#1}}

\newcommand{\gf}[1]{\textcolor{black}{#1}}
\newcommand{\ph}[1]{\textcolor{black}{#1}}

\newcommand{\hd}[1]{\textcolor{black}{#1}}

\IEEEoverridecommandlockouts 

\begin{document}
\date{}

\title{5G New Radio for Automotive, Rail, and Air Transport} 
\singlespacing

\author{
G\'{a}bor Fodor$^{\star\ddag}$,
Julia Vinogradova$^{\dag}$,
Peter Hammarberg$^{\star}$,
Keerthi Kumar Nagalapur$^{\star}$,\\
Zhiqiang (Tyler) Qi$^\flat$,
Hieu Do$^{\star}$,
Ricardo Blasco$^{\dag}$,
Mirza Uzair Baig$^{\star}$
\\

\small $^\star$Ericsson Research, Sweden, E-mail: \texttt{firstname.secondname@ericsson.com} \\
\small $^\ddag$KTH Royal Institute of Technology, Sweden. E-mail: \texttt{gaborf@kth.se}\\
\small $^\dag$Ericsson Research, Finland, Email: \texttt{Julia.Vinogradova|Ricardo.Blasco@ericsson.com}\\
\small $^\flat$Ericsson Research, China, Email: \texttt{Zhiqiang.Qi@ericsson.com}
}


\begin{acronym}[LTE-Advanced]
  \acro{2G}{Second Generation}
  \acro{3G}{3$^\text{rd}$~Generation}
  \acro{3GPP}{3$\text{rd}$~Generation Partnership Project}
  \acro{4G}{4$^\text{th}$~Generation}
  \acro{5G}{5$^\text{th}$~Generation}
  \acro{5GPPP}{5G Infrastructure Public Private Partnership}
  \acro{QAM}{quadrature amplitude modulation}
  \acro{ADAS}{Advanced driver assistance system}
  \acro{AD}{autonomous driving}
  \acro{AI}{artificial intelligence}
  \acro{AoA}{angle of arrival}
  \acro{AoD}{angle of departure}
  \acro{API}{application programming interface}
  \acro{AR}{autoregressive}
  \acro{ARQ}{automatic repeat request}
  \acro{BER}{bit error rate}
  \acro{BLER}{block error rate}
  \acro{BPC}{Binary Power Control}
  \acro{BPSK}{Binary Phase-Shift Keying}
  \acro{BRA}{Balanced Random Allocation}
  \acro{BS}{base station}
  \acro{CAM}{cooperative awareness messages}
  \acro{CAP}{Combinatorial Allocation Problem}
  \acro{CAPEX}{capital expenditure}
  \acro{CBF}{coordinated beamforming}
  \acro{CBR}{congestion busy ratio}
  \acro{CDD}{cyclic delay diversity}
  \acro{CDF}{cumulative distribution function}
  \acro{CDL}{clustered delay line}
  \acro{CS}{Coordinated Scheduling}
  \acro{C-ITS}{cooperative intelligent transportation system}
  \acro{CSI}{channel state information}
  \acro{CSIT}{channel state information at the transmitter}
  \acro{D2D}{device-to-device}
  \acro{DCA}{Dynamic Channel Allocation}
  \acro{DCI}{downlink control information}
  \acro{DE}{Differential Evolution}
  \acro{DENM}{decentralized environmental notification messages}
  \acro{DFO}{Doppler frequency offset}
  \acro{DFT}{Discrete Fourier Transform}
  \acro{DIST}{Distance}
  \acro{DL}{downlink}
  \acro{DMA}{Double Moving Average}
  \acro{DMRS}{Demodulation Reference Signal}
  \acro{D2DM}{D2D Mode}
  \acro{DMS}{D2D Mode Selection}
  \acro{DMRS}{demodulation reference symbol}
  \acro{DPC}{Dirty paper coding}
  \acro{DPS}{Dynamic point switching}
  \acro{DRA}{Dynamic resource assignment}
  \acro{DSA}{Dynamic spectrum access}
  \acro{eMBB}{enhanced mobile broadband}
  \acro{eV2X}{Enhanced vehicle-to-everything}
  \acro{EIRP}{equivalent isotropically radiated power}
  \acro{ERTMS}{European Rail Traffic Management System}
  \acro{ETSI}{European Telecommunications Standards Institute}
  \acro{FDD}{frequency division duplexing}
  \acro{FR1}{frequency range-1}
  \acro{FR2}{frequency range-2}
  \acro{GNSS}{global navigation satellite system}
  \acro{HARQ}{hybrid automatic repeat request}
  \acro{HST}{high-speed train}
  \acro{IAB}{integrated access and backhaul}
  \acro{ITS}{intelligent transportation system}
  \acro{KPI}{key performance indicator}
  \acro{IEEE}{Institute of Electronics and Electrical Engineers}
  \acro{IMT}{International Mobile Telecommunications}
  \acro{IMU}{inertial measurement unit}
  \acro{InC}{in-coverage}
  \acro{IoT}{Internet of Things}
  \acro{ITS}{intelligent transportation system}
  \acro{LDPC}{low-density parity-check coding}
  \acro{LMR}{land mobile radio}
  \acro{LoS}{line-of-sight}
  \acro{LTE}{Long Term Evolution}
  \acro{MAC}{medium access control}
  \acro{mmWave}{millimeter-wave}
  \acro{MBB}{mobile broadband}
  \acro{MCS}{modulation and coding scheme}
  \acro{METIS}{Mobile Enablers for the Twenty-Twenty Information Society}
  \acro{MIMO}{multiple-input multiple-output}
  \acro{MISO}{multiple-input single-output}
  \acro{ML}{machine learning}
  \acro{MRC}{maximum ratio combining}
  \acro{MS}{mode selection}
  \acro{MSE}{mean square error}
  \acro{MTC}{machine type communications}
  \acro{multi-TRP}{multiple transmission and reception points}
  \acro{mMTC}{massive machine type communications}
  \acro{cMTC}{critical machine type communications}
  \acro{NDAF}{Network Data Analytics Function}
  \acro{NF}{network function}
  \acro{NR}{New Radio}
  \acro{NSPS}{national security and public safety}
  \acro{NWC}{network coding}
  \acro{OEM}{original equipment manufacturer}
  \acro{OFDM}{orthogonal frequency division multiplexing}
  \acro{OoC}{out-of-coverage}
  \acro{PSBCH}{physical sidelink broadcast channel}
  \acro{PSFCH}{physical sidelink feedback channel}
  \acro{PSCCH}{physical sidelink control channel}
  \acro{PSSCH}{physical sidelink shared channel}
  \acro{PDCCH}{physical downlink control channel}
  \acro{PDCP}{packet data convergence protocol}
  \acro{PHY}{physical}
  \acro{PLNC}{physical layer network coding}
  \acro{PPPP}{proximity services per packet priority}
  \acro{PPPR}{proximity services per packet reliability}
  \acro{PSD}{power spectral density}
  \acro{RLC}{radio link control}
  \acro{QAM}{quadrature amplitude modulation}
  \acro{QCL}{quasi co-location}
  \acro{QoS}{quality of service}
  \acro{QPSK}{quadrature phase shift keying}
  \acro{PaC}{partial coverage}
  \acro{RAISES}{Reallocation-based Assignment for Improved Spectral Efficiency and Satisfaction}
  \acro{RAN}{radio access network}
  \acro{RA}{Resource Allocation}
  \acro{RAT}{Radio Access Technology}
  \acro{RB}{resource block}
  \acro{RF}{radio frequency}
  \acro{RS}{reference signal}
  \acro{RSRP}{Reference Signal Received Power}
  \acro{SA}{scheduling assignment}
  \acro{SFN}{Single frequency network}
  \acro{SNR}{signal-to-noise ratio}
  \acro{SINR}{signal-to-interference-plus-noise ratio}
  \acro{SC-FDM}{single carrier frequency division modulation}
  \acro{SFBC}{space-frequency block coding}
  \acro{SCI}{sidelink control information}
  \acro{SL}{sidelink}
  \acro{SLAM}{simultaneous localization and mapping}
	\acro{SPS}{semi-persistent scheduling}
  \acro{STC}{space-time coding}
  \acro{SW}{software}
  \acro{TCI}{transmission configuration indication}
  \acro{TBS}{transmission block size}
  \acro{TDD}{time division duplexing}
  \acro{TRP}{transmission and reception point}
  \acro{TTI}{transmission time interval}
  \acro{UAV}{unmanned aerial vehicle}
  \acro{UAM}{urban air mobility}
  \acro{UE}{user equipment}
  \acro{UL}{uplink}
  \acro{URLLC}{ultra-reliable and low latency communications}
  \acro{VUE}{vehicular user equipment}
  \acro{V2I}{vehicle-to-infrastructure}
  \acro{V2N}{vehicle-to-network}
  \acro{V2X}{vehicle-to-everything}
  \acro{V2V}{vehicle-to-vehicle}
  \acro{V2P}{vehicle-to-pedestrian}
  \acro{ZF}{Zero-Forcing}
  \acro{ZMCSCG}{Zero Mean Circularly Symmetric Complex Gaussian}
 \acro{TBS}{transport block size}
 \acro{SCI}{sidelink control information}
\end{acronym}

\maketitle
\pagestyle{empty}
\thispagestyle{empty}

\begin{abstract}
The recent and upcoming releases of the 3rd Generation Partnership Project's 5G New Radio (NR) specifications include
features that are motivated by providing connectivity services to a broad set of verticals, including
the automotive, rail, and air transport industries.
Currently, several radio access network features are being further enhanced or newly introduced
in NR to improve 5G's capability to provide fast, reliable, and non-limiting connectivity for transport applications.
In this article, we review the most important
characteristics and requirements of a wide range of services that are driven by the desire to help the transport sector to
become more sustainable, economically viable, safe, and secure.
These requirements will be supported
by the evolving and entirely new features of 5G NR systems, including
accurate positioning,
\gf{reference signal design to enable multi-transmission and reception points},
\gf{service-specific} scheduling configuration,
and service quality prediction.


\end{abstract}

\section{Introduction}
\label{Sec:Intro}

Recent advances in wireless communications, real-time control, sensing, and
battery technologies, collaborative spectrum management and sharing,
and artificial intelligence are enabling the transport sector to become
more cost efficient, secure, and sustainable \cite{Soriano:18}. 
Due to new requirements arising in road, railway, air and maritime transport, cellular connectivity,
and reliable wireless communications between vehicles and road users are no longer a "nice to have",
but are essential parts of \acp{C-ITS} 
and smart cities \cite{Zeadally:20}. 
Ericsson predicts that the number of connected cars in operation will rise to more than 500 million in 2025,
and the railway sector is making the first steps to digitalize the \ac{ERTMS}, which includes
mission-critical control systems for train operations, including \acp{HST}.
The \ac{UAV} and \ac{UAM} (drone) market is expected to grow from the current estimated USD $4.4$bn to $63.6$bn by 2025
\cite{Alwateer:20}.
Apart from smart city applications, there is a growing interest in employing connected \acp{UAV} in
\gf{surface} mining, seaports, oil and gas, and other large industrial facilities or in public safety situations in order to improve and
optimize industrial processes, enhance operational efficiencies, and create resilience. 

The digitalization and increasing connectivity in the transport sector are driven by three key factors.
First, there are increasing demands imposed virtually by all stakeholders -- including passengers,
cargo companies, vehicle (car, truck, locomotive, ship) manufacturers, public transport and rail operators,
and infrastructure (road, rail, harbor) providers.
This broad set of requirements includes being always connected to the Internet and enterprise
networks, enjoying safe and secure journeys in urban and rural environments, and minimizing
environmental impacts.
At the same time, reducing capital and operational expenditures necessitates increasing digitalization,
automation, and always-on connectivity, since these technologies make manufacturing, maintaining and operating
transport equipment, infrastructure, and services much more efficient.
Thirdly, the rapid deployment of 5G networks, and the recent advances in 6G research provide a technology
push towards digitalized and connected transport services \cite{Samdanis:20}.

\gf{In parallel with} the above trends in the transport industry, 
the release 15 (Rel-15) of the
\ac{3GPP} specifications in 2016 marked the birth of the new cellular radio interface for
\hd{the} fifth generation (5G) systems,
commonly referred to as \ac{NR}.
Although \ac{MBB} services continue to be the main driver for \ac{NR},
\hd{this new radio technology generation inherently has much stronger support for verticals such as the transport industry, as compared to \ac{LTE}. Additionally, already in Rel-16,
new technical features are introduced specifically for supporting critical machine-type communications including \ac{URLLC} \ac{V2X} services for automotive.
Further enhancements targeting special connectivity requirements of the rail operations and remote control of \acp{UAV} are also being standardized in the upcoming releases.}

Compared with 4G systems, \gf{5G \ac{NR}}
\hd{adopts a new design philosophy and novel technology components,}
including flexible numerology and waveform design for lower
and millimeter-wave frequency bands, minimizing control signaling overhead,
multi-hop support by integrated access and backhaul relay, 
enhanced positioning, and \ac{QoS} handling mechanisms.
Also, further enhanced \ac{MIMO} techniques enable 5G networks to acquire accurate \ac{CSI} for both
analog and hybrid beamforming and spatial multiplexing applications, which are important
for maintaining high spectral efficiency even in high-speed road and rail transport scenarios \cite{Han:19}.
Finally, recent 3GPP releases of 5G radio access networks pave the way for advanced \hd{radio-based} positioning techniques
that efficiently complement and make positioning more precise than pure satellite-based positioning
techniques \cite{Lin:17}, 
which are highly useful for
\hd{automotive and drone use cases.}

\hd{The present paper serves two purposes. Firstly, we summarize the technical foundations of 5G NR which can fulfill basic requirements imposed by emerging
use cases in the transport sector. Secondly, based on an in-depth review of connectivity requirements of transport use cases, we highlight several important
new technology enablers which will play a key role in meeting \gf{the most} stringent requirements.
In particular, we focus on the following,}
\begin{itemize}
\item
Positioning techniques that take advantage of \gf{combining onboard sensors} and cellular network measurements;
\item
Reference signal design and selecting the appropriate multi-\ac{TRP} option
for spectrum-efficient operations of \acp{HST} and other high-speed \acp{UE};
\item
Service-specific scheduling techniques for \ac{V2X} communications that ensure high resource utilization and
service differentiation between low-latency and delay tolerant (lower than best effort) traffic types;
\item
Novel \ac{QoS}-prediction techniques that are useful in \gf{driverless and driver-assisted} road, rail, and drone transport use cases.
\end{itemize}
\hd{\section{Technical Foundations of 5G NR and the Initial NR Evolution Targeting the Transport Vertical}}
\begin{figure}[t]
\begin{center}
\includegraphics[width=1.05\columnwidth]{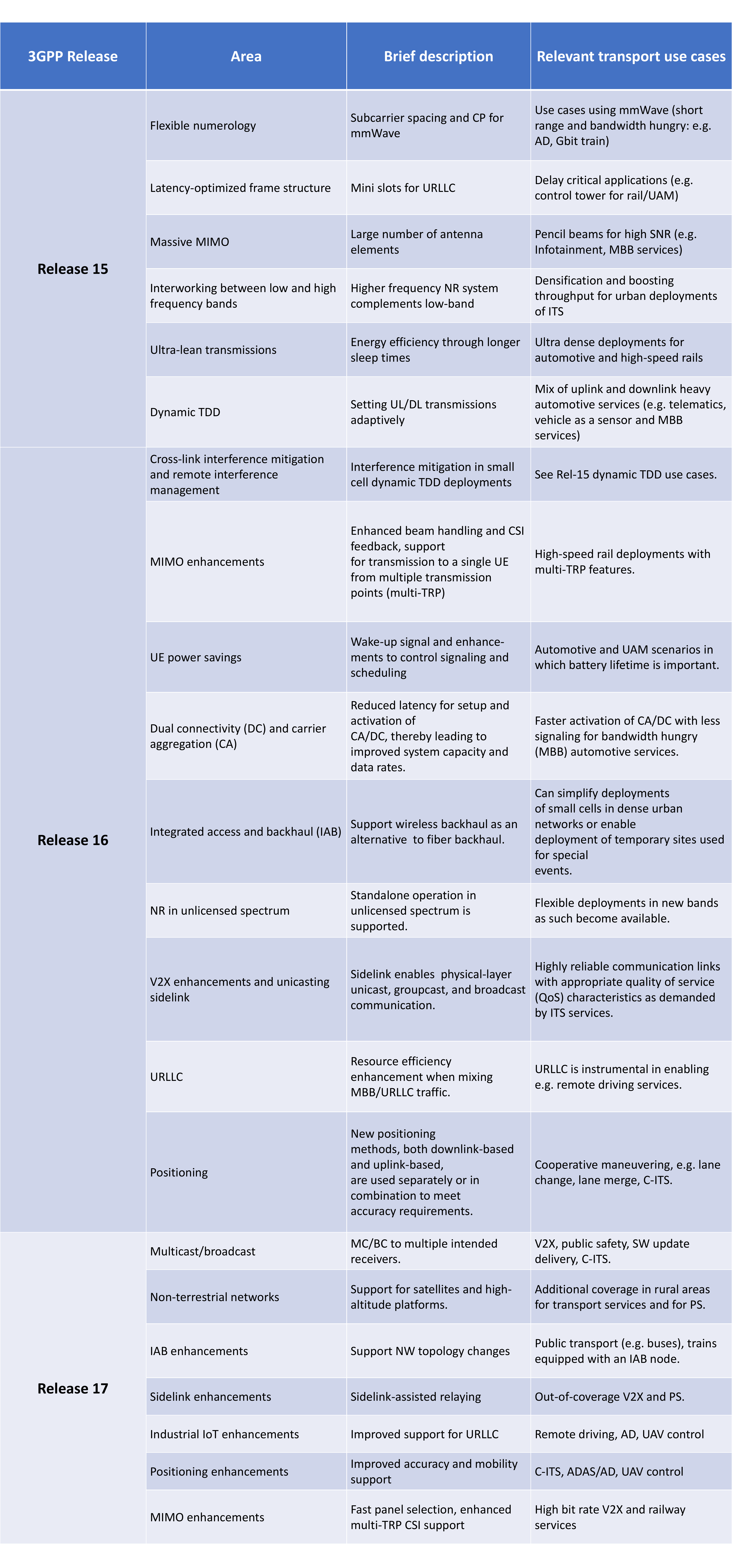}
\caption{Key areas and relevant transport use cases in NR Rel-15, Rel-16, and Rel-17.}
\label{Fig:Table}
\end{center}
\end{figure}

\subsection{Major Features in Rel-15 and Rel-16}
As mentioned, Rel-15 and Rel-16 of the \ac{3GPP} specifications have been largely driven by requirements of
\ac{MBB} services, including requirements on enhanced data rates, latency, coverage, capacity, and reliability. 
However, starting already in Rel-15 and continuing in the subsequent \ac{3GPP} releases, \ac{NR} enables new use cases by
meeting the requirements imposed by transport use cases, such as connected cars, high-speed trains, and \acp{UAV}.
While Rel-15 focused on supporting \ac{MBB} and \ac{URLLC} applications, Rel-16/17 includes UE power savings,
operation in unlicensed spectrum, industrial \ac{IoT} enhancements
as well as special \ac{RAN} features such as physical layer support for unicasting sidelink (device-to-device) for advanced \ac{V2X}
services \cite{Ashraf:20}.

A key distinguishing feature of 5G \ac{NR} from fourth generation (4G) systems is the substantial expansion of the frequency bands,
in which \ac{NR} can be deployed.
For transport applications,
the following \ac{NR}-specific features are particularly important (see Figure \ref{Fig:Table}):
\begin{itemize}
\item
Symmetric physical layer design with \ac{OFDM} waveform for all link types, including uplink, downlink, sidelink,
and backhaul;
\item
Wide range of carrier frequencies, bandwidths, and deployment options. 3GPP aims to develop and specify components
and physical layer numerology that can operate in frequencies up to 100 GHz. This implies several options for
\ac{OFDM} subcarrier spacing ranging from 15 kHz up to 240 kHz with a proportional change in cyclic prefix duration;
\item
Native support for dynamic \ac{TDD} as a key technology component.
In dynamic \ac{TDD}, parts of a slot can be
adaptively allocated to either uplink or downlink, depending on the prevailing traffic demands;
\item
Support for massive \ac{MIMO}, that is a massive number of steerable antenna elements for both transmission and
reception, utilizing channel reciprocity in \ac{TDD} deployments and a flexible \ac{CSI} acquisition framework.
\gf{The} NR channels and signals, including those used for data transmission, control signaling and synchronization
are all designed for optional beamforming.
\end{itemize}

In addition to flexible numerology, native support for dynamic \ac{TDD} and advanced \ac{MIMO} features,
\ac{NR} is designed using the principle of ultra-lean design, 
which aims at minimizing
control plane and synchronization signal transmissions when data transmissions are idle.
Inherent support for distributed \ac{MIMO}, also referred to as multi-\ac{TRP}, is introduced
and fully supported in Rel-16.
This feature is largely motivated by \ac{HST} applications, since it
allows \acp{UE} to receive multiple control and data channels
per slot, which enables simultaneous data transmissions from multiple physically separated base stations.

\subsection{Major Developments in Rel-17}
Looking beyond Rel-16, the NR evolution will be driven by
industry verticals, including a variety of transport use cases,
such as \ac{V2X} communications, high-speed trains, \acp{UAV} and passenger aircrafts, and maritime
communications.
These use cases justify new features discussed and planned for Rel-17.
\ac{MIMO} enhancements
are expected to support multi-\ac{TRP} specific tracking reference signals, single frequency network deployments,
and non-coherent joint transmissions by multiple base stations, which are particularly useful for providing
connectivity to high-speed trains.
Furthermore,
Rel-17 is studying the integration of non-terrestrial and terrestrial networks in order to support
use cases for which terrestrial networks alone cannot provide the required coverage and capacity, including maritime,
\ac{UAV}, and \ac{UAM} 
scenarios.

\section{Overview of Intelligent Transportation Systems 
Services and Requirements}


\begin{figure}[t]
\begin{center}
\includegraphics[width=1\columnwidth]{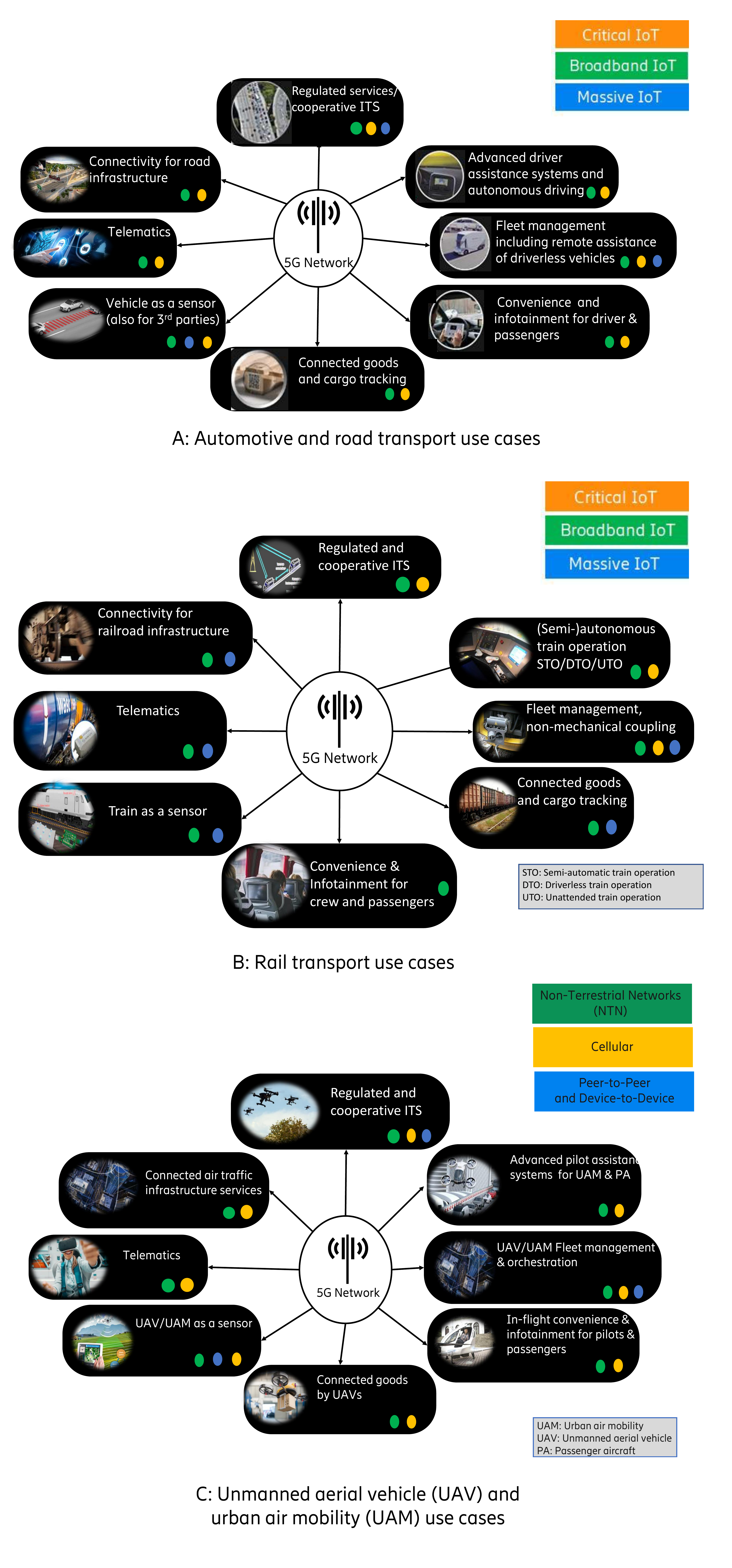}
\caption{Use case \gf{categories} in the automotive and road transport (upper), railway (middle), and \ac{UAV} and \ac{UAM} (lower) segments.}
\label{Fig:Flower}
\end{center}
\end{figure}

Figure \ref{Fig:Flower} classifies the automotive, rail, and \ac{UAV}/\ac{UAM} use cases in use case categories,
\hd{together with the key connectivity requirements per category.}

Regulated \acp{C-ITS} 
provide
international or governmental regulated services for road, rail and drone traffic efficiency, sustainability, and safety.
Traffic efficiency use cases have relaxed latency requirements, while safety-related data often
requires \ac{URLLC}.
A benefit of regulation is to facilitate 
\ac{OEM} cooperation in standardized information exchange.
\ac{C-ITS} services may also use dedicated 
spectrum in certain regions; for example,
for direct short-range communication using the 3GPP 
\gf{sidelink} 
technology.
For rail transport, \ac{C-ITS} implies
station dwell time control and speed/break control to optimize rail network utilization while ensuring safety.

\acp{ADAS} and \ac{AD} are increasingly employed for both road and rail
transport.
In Europe, for example, the next generation of the \ac{ERTMS} will
support well-defined levels of automation, including semi-automated (assisted) driving, driverless
and unattended train operation. 
Similarly, advanced pilot assistance systems for \ac{UAM} and passenger aircrafts are envisioned by various
stake-holders of the air transport industry. 
\hd{For this set of applications, \ac{URLLC} communication and high-accuracy positioning play crucial roles.}

Fleet management including remote assistance of driverless vehicles is an important application for
road, rail, and \ac{UAV}-based transport.
\hd{This type of} services aim at vehicle fleet owners such
as logistics or car-sharing companies. The communication service is primarily used to monitor vehicle
locations and the vehicle/driver status. With increasing level of automation in the rail industry and
for \acp{UAV}, or for a fleet of driverless trucks, fleet management also includes communication support
for operations monitoring and remote assistance from a control tower. 

Convenience and infotainment, based on \ac{MBB} services for drivers, crew, and passengers are
equally important in road, rail, and future \ac{UAM} transport use cases. Such services deliver
content such as traffic news and audio entertainment for car drivers, and gaming and video entertainment for
passengers.
\hd{One specific example is the concept of "Gigabit train" services, which motivate the adoption of \ac{HST} scenarios in 3GPP. For this set of use cases, the most important requirement is high data rate and low latency connections, which rely heavily on the capability of tracking wireless channels at \gf{high vehicle speed}.}

The primary focus in the logistics and connected goods category is on the tracking of transported
objects (commodities, merchandise goods, cargo and so on) during the production and transport
cycle of the object. Near real time tracking and status monitoring of goods are attractive for cargo
companies, customers, and freight train operators.

In the \hd{vehicle-as-a-sensor} use case category, sensors installed in the vehicle
sense the environment and can also provide anonymized data to 3rd parties.
In road transport, for example,
\hd{the vehicle-mounted sensors}
provide information
\hd{for solutions such as \ac{ADAS} or \ac{AD} as well as for monitoring city infrastructure and road
status.}
For rails, railway track monitoring and anomaly detection are supported by various sensors mounted on the train,
effectively operating the train as a sensor. 
\hd{Similarly, drones can be equipped with a lot of sensors that help collect data and perform distributed or federated computation for various purposes such as forecasting cloud formation, rain, and other hazardous weather conditions.
Just as with the convenience and infotaiment \gf{category}, the vehicle-as-a-sensor requires high data-rate connections between vehicles or between vehicles and the cellular network at high vehicle speed and dynamic interference conditions.}

Telematics applications for vehicles include collecting vehicle diagnostics to monitor/adjust the vehicle,
while rail telematics rail applications allow continuous status updates from trains to determine state, delay,
cargo conditions, \ac{SW} updates, and geo-fencing.
\gf{In this category, several applications (e.g. \ac{SW} updates) tolerate delay, while others are more delay critical.}
Similarly, for air transport, telematics serve as a tool for
collecting air vehicles diagnostics to monitor/adjust the vehicles.
\hd{To make sense of the vast amount of data collected from vehicles in this group of use cases, the role of \ac{AI}/\ac{ML} is utterly important.
In the reverse direction, \ac{AI}/\ac{ML} can also play a meaningful role in determining
when to perform a certain task to the vehicle in an efficient manner.
For example, \ac{ML}-based spare capacity prediction, which is part of the so-called cellular network \ac{QoS} prediction,
can be used to predict the most economical time for \ac{SW} update for a set of vehicles \cite{5GAA:19, Raca:20}.}

Connected road infrastructure services are operated by cities and road authorities to monitor
the state of the traffic and control its flow, such as physical traffic guidance systems, parking
management and dynamic traffic signs. For railways,
\hd{the communication between the rail infrastructure and the locomotive via specific transmission modules and eurobalises is used to send information from line-side electric units to the trains
e.g., current speed restrictions for the coming rail segment.
For UAV/UAM, an Unmanned Aircraft System Traffic Management (UTM) is used for traffic control,
which requires high-accuracy 3D positioning and \ac{URLLC} communication with the ground control system.}

\section{Proposed New Features and Solutions to Support \ac{ITS} Requirements}

\hd{Despite the recent and ongoing enhancements to 5G NR, there is still a need
to further improve the technology to meet the growing demands of industry verticals.
In this section, we summarize the state of the standardization of several specific features
and introduce new solutions which can help fulfilling the stringent connectivity requirements
of the transport sectors outlined in the preceding section.
\gf{These components span both radio layers (physical, medium access control) and the service layer of the protocol stack.}
}

\subsection{Advanced Positioning Support and Algorithms}

With the introduction of \ac{NR}, \ac{3GPP} targets improved positioning capabilities to cater
for a number of new use cases, involving indoor, industrial, and automotive applications.
Cooperative manoeuvring in the
\gf{\ac{C-ITS} category and several \ac{ADAS} applications
rely on accurate positioning, which must remain operational even in \ac{GNSS}-problematic areas.}
Specifically in Rel-16, a set of positioning related features are introduced,
which pave the way for enhanced positioning services.
These new features include new and improved \ac{UL} and \ac{DL} reference signal design, allowing larger bandwidths and beamforming,
assisted by additional measurements and enhanced reporting capabilities.
By supporting larger bandwidths than in \ac{LTE}, higher accuracy of range estimates can be achieved,
and with \ac{AoA} and \ac{AoD} 
measurements new positioning schemes exploiting the spatial domain can be supported.

Architecture-wise, similarly to \ac{LTE}, \ac{NR} positioning is based on the use of a location server.
The location server collects and distributes information related to positioning
(UE capabilities, assistance data, measurements, position estimates, etc.) to the other entities
involved in the positioning procedures (\acp{BS} and connected vehicles). A range of positioning
schemes, including \ac{DL}-based and \ac{UL}-based ones, are used separately or in combination to meet the
accuracy requirements in vehicular scenarios.

Specifically, in the \ac{mmWave}-bands, referred to as \ac{FR2} bands of \ac{NR},
the \ac{AoA} and \ac{AoD} measurements can be enhanced
by using 
large antenna arrays, which facilitate high resolution angular measurements. 
By unlocking the spatial domain, \ac{NR} can significantly increase the positioning accuracies
for many industrial and automotive use cases \cite{Wymeersch:17}, \cite{Mostafavi:20}.
Additionally, \gf{to further improve the accuracy and reliability of positioning in \ac{GNSS}-problematic areas, we propose to fuse}
acceleration measurements provided by onboard \acp{IMU} 
with measurements on the received \ac{NR} \ac{DL} reference signals \gf{from multiple \ac{NR} \acp{BS}}.
Fusing local \ac{IMU} measurements with measurements on \gf{multiple \ac{NR} \ac{DL} signals} can 
reach decimeter accuracy in favourable deployments. 

Furthermore, high accuracy spatial and temporal measurements facilitate
the use of advanced positioning schemes such as \ac{SLAM}, 
which utilizes consecutive measurements to build a statistical model of the environment, 
achieving high accuracy
even in extreme scenarios by utilizing measurements on \ac{DL} signals of only a single base station. 

\subsection{Further Enhanced \ac{MIMO} to Support Multiple Transmission and Reception Points}

\ac{HST} wireless communication is characterized by a highly time-varying channel
and rapid changes of the closest \acp{TRP} to the train, 
resulting from the extreme high velocities.
Recognizing these challenges, \ac{NR} has been designed to support high mobility from day one, 
and includes features to enable communications with \acp{HST} \cite{Noh:20}.
Furthermore, several multi-\ac{TRP} deployment options and features developed 
under the general \ac{MIMO} framework can be exploited to support \ac{HST} communications \gf{for the "Gigabit train"},
while minimizing the need for handovers. 
\gf{We expect this technology component will also play a very important role in \ac{UAV}/\ac{UAM} use cases.}

The multi-\ac{TRP} options that are the most relevant to \ac{HST} communications are:
\begin{itemize}
\item
\ac{DPS}:
Data signals are transmitted from a single \ac{TRP} at a given time, and the \ac{TRP}
used for transmission is dynamically selected based on the relative quality of channels between the train and a few closest \acp{TRP};
\item
\ac{SFN}:
All the \acp{TRP} in the \ac{SFN} area transmit the same data and reference signals to the train;
\item
\ac{SFN} with \ac{TRP}-specific reference signals:
The same data signal is transmitted from different \acp{TRP}, while some of the reference signals are transmitted
in a \ac{TRP}-specific manner.
\end{itemize}

\hd{The first and the third options}
rely on \ac{TRP}-specific reference signals, whereas
\hd{the second option} uses common reference signal
across the \acp{TRP} in the coverage area of the \ac{SFN}.
In addition to supporting \ac{TRP}-specific reference signals,
\ac{NR} supports associating different reference signals with different channel properties,
such as Doppler shift and delay spread, through \ac{NR}'s \ac{QCL} and \gf{transmission configuration indication} framework. 
In Rel-17, different \ac{QCL} enhancements are investigated to better support
advanced channel estimation schemes that can be implemented at the train
and to evaluate the need for \ac{TRP}-side pre-compensation algorithms. 
\gf{In Section~\ref{Sec:Num}, we evaluate the necessity and performance of the multi-TRP options described above
through link level evaluations.}
Furthermore, beam management enhancements necessary to support \ac{HST} communications
in the higher bands of \ac{NR} are also investigated in Rel-17.
\subsection{Service-Specific Scheduling}
To accommodate \ac{MBB}, \ac{URLLC}, and machine-type communications 
services, \ac{NR} networks employ scheduling algorithms
that take into account the current service mix, prevailing channel conditions, traffic load,
available carriers, and other factors.
Scheduling in multiservice wireless networks has been researched
for decades and a vast literature as well as practically deployed scheduling schemes exist.
Interestingly, due to requirements imposed by the coexistence of \ac{URLLC} and delay tolerant services,
\ac{mmWave} communications support in the \ac{FR2} bands and serving very high speed \ac{UE}-s
have stimulated renewed research interest in this topic \cite{Esswie:18}. 
%

Some recent works propose scheduling strategies to minimize end-to-end delay for time-critical
services \cite{Hu:17}, or to optimize resource allocation for the coexistence of various services \cite{Esswie:18}.
In \ac{NR}, service-specific scheduling can also be configured to take into account the specific opportunities
that are present in certain deployment scenarios. 
\gf{We propose to
customize} the scheduler in certain deployment scenarios, which can be illustrated by
a scheduling configuration that is suitable for
non-time-critical services.
\gf{This can be applicable
for background data transfer for \ac{SW} updates or uploading sensor measurements in the vehicle-as-a-sensor category.}
This scheduling mechanism divides vehicles into a high and a low path-gain group.
Vehicles belonging to the high path-gain group are eligible for medium access, while scheduling vehicles
in the low path-gain group is postponed (dropped) until their path gain improves.
The scheme is suitable for highly mobile users (including automotive or urban train use cases)
in high-way, urban or suburban scenarios, and can be activated or de-activated based on velocity or
other sensor measurements that help to configure the path gain threshold.

\subsection{QoS Prediction} 
The NR \ac{QoS} framework together with features like \ac{URLLC} are successful in delivering
a minimum guaranteed performance, especially in controlled scenarios.
However, highly mobile \acp{UE} usually experience time-variant network performance,
partly because the actual \ac{QoS} often exceeds the minimum or guaranteed level,
and partly because the system is occasionally not able to fulfill a \ac{QoS} provision.

Interestingly, in many cases, \gf{including certain \ac{C-ITS}, \ac{ADAS} or telematics applications,}
these performance fluctuations are not a problem if
they can be predicted in advance. Having access to real-time \ac{QoS} predictions
has generated a large interest from the automotive industry \cite{5GAA:19},
as it would allow service providers, mobile network users, and automotive applications to dynamically adapt their behaviors
to the prevailing or imminent \ac{QoS} level.
This would allow for enabling services relying on continuous guaranteed performance
as well as for exploiting spare capacity for large bulk data transfers in lower than best effort services.

Despite the high expectations, \ac{QoS} prediction is largely an open research topic.
The realistically achievable performance and the applicability of this type of algorithms are still unknown.
To a large extent, \ac{QoS} prediction is seen as a \ac{ML} application
with a broad data set consisting of network measurements, device measurements,
and application data \cite{Raca:20}.

In practice, different types of information may be collected with different periodicities,
time horizons, resolutions, and accuracies, depending on practical and business-related constraints.
Understanding the relevance of each of them is ongoing work and will be instrumental
in determining the relative merits and tradeoffs of the different architecture options,
which range from network-centric to application-based.
As a first step towards supporting predictive \ac{QoS} in mobile networks,
3GPP enhanced the \ac{NR} system architecture in Rel-16 to support services
providing network data analytics in the 5G core network.
To this end, \gf{application programming interfaces} for exposing network-based predictions were defined,
the necessary procedures for collecting the relevant data for the analytics from different
\gf{network functions} as well as from the operations administration and management functionality. 

In addition, procedures providing analytics
(e.g., load, network performance, data congestion, \ac{QoS} sustainability and \ac{UE} related analytics)
to other \gf{network functions} were introduced. As usual, the algorithms used to obtain the network analytics
are not defined in the specification, and as said, they are a topic of ongoing research.

\section{Numerical Examples}
\label{Sec:Num}

\subsection{Positioning}

\begin{figure}[t]
\begin{center}
\includegraphics[width=0.95\columnwidth]{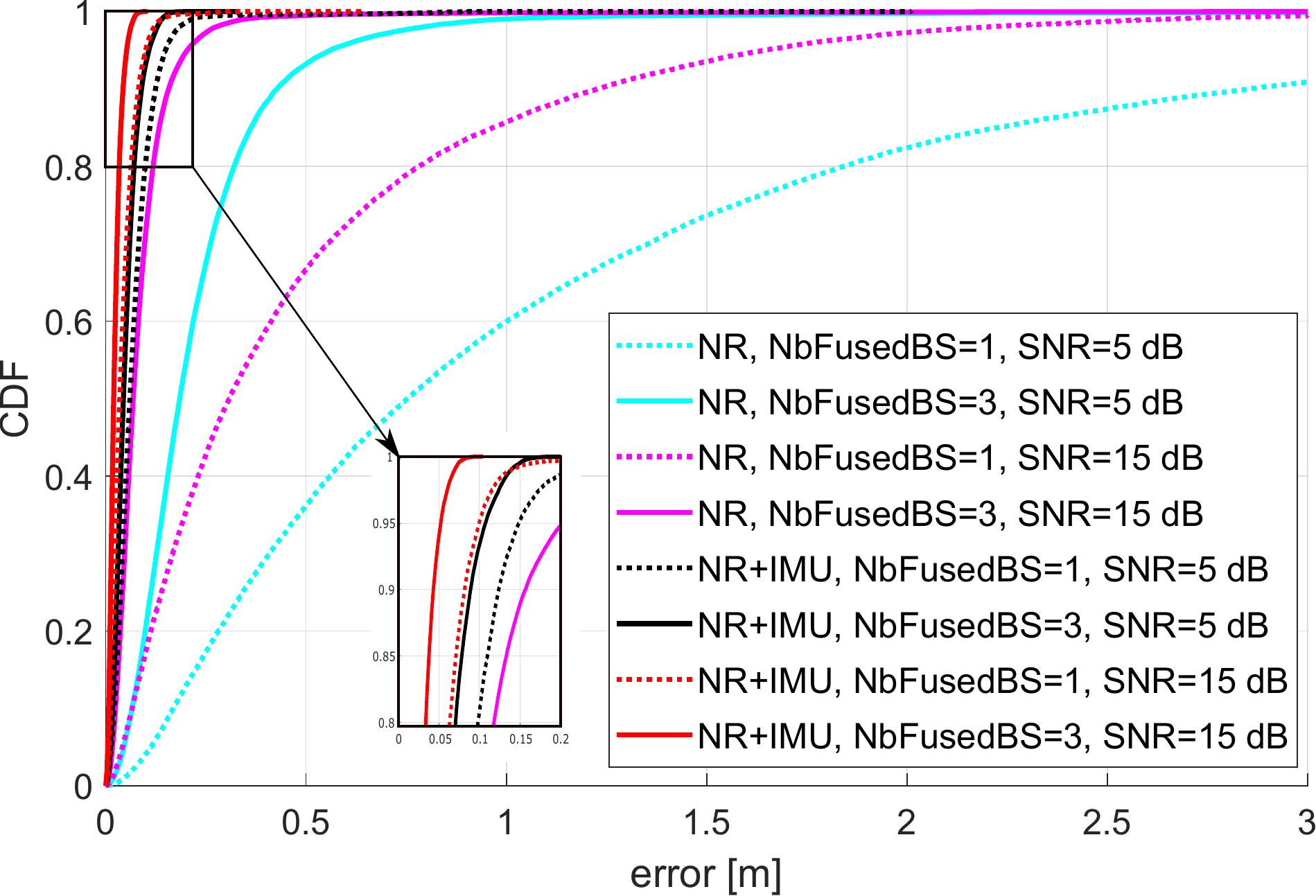}
\caption{CDF of the obtained positioning accuracy when using 
NR only and fused IMU+NR positioning at 5 and 15 dB SNR.}
\label{Fig:Pos}
\end{center}
\end{figure}

We consider a highway scenario with \acp{BS} equally spaced
and placed at the side of the road. A moving vehicle following
a snake-like trajectory is equipped with an onboard \ac{IMU} sensor
and an \ac{NR} \ac{UE}.
A \ac{MIMO} system is considered with square antenna arrays at the \ac{BS} and the \ac{UE}.
A \ac{LoS} downlink propagation is assumed with a grid of \gf{Discrete Fourier Transform} beams transmitted by the \ac{BS}.
A sensor fusion-based positioning approach is proposed (similar to the one in \cite{Mostafavi:20}),
for which a Kalman filter is used to fuse the measurements obtained from the \ac{IMU} and the \ac{NR} downlink
such as the range and the angles-of-arrival.
An extension to this is proposed allowing to fuse measurements from multiple \ac{BS}.

Positioning accuracy in terms of positioning \acp{CDF} are compared in Figure \ref{Fig:Pos}
for \ac{NR}-based method and for the sensor fusion-based method combining \ac{IMU} with \ac{NR}.
It is assumed that the \acp{BS} operate at the millimeter wave frequency of 28 GHz with 256 antennas.
There are 4 antennas at the UE, the UE speed \gf{is} equal to 130 km/h, and \ac{SNR} \gf{is} equal to 5 and 15 dB.
The \acp{BS} are placed at 40 m from the road with the inter-site distance equal to 200 m.
The results are averaged over a total distance of 10 km.
The number of \acp{BS} used to fuse the measurements from is denoted by NbFusedBS.
The simulation results show that a large performance gain is obtained for the sensor
fusion-based method as compared to the \ac{NR} only-based method, especially \gf{at} low \ac{SNR}.
\gf{Notice that fusion-based methods allow to achieve a decimeter level accuracy
with greater than 90\% probability, even at low \ac{SNR} of 5dB.}

\subsection{Further Enhanced \ac{MIMO} to Support Multiple Transmission and Reception Points}

\begin{figure}[t]
\begin{center}
\includegraphics[width=1.1\columnwidth]{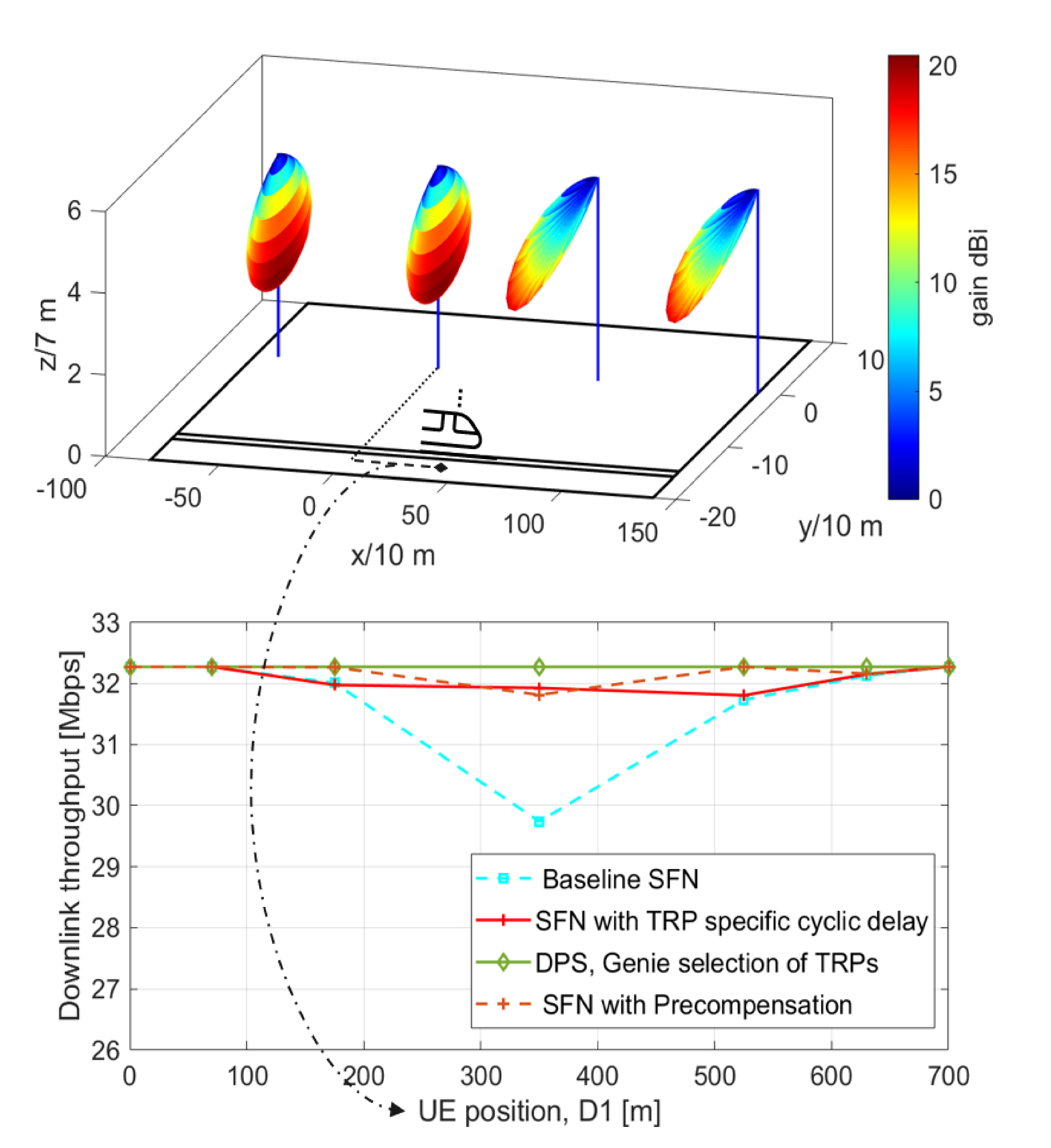}
\caption{High-speed train multi-\ac{TRP} scenario with four base stations (upper) and downlink throughput
as a function of train position when using \ac{SFN} without/with \ac{CDD}, \ac{DPS} and \ac{SFN} with precompensation.
(The $x, y$ and $z$ axes are scaled by 10, 10 and 7 respectively for better visualization.)
}
\label{Fig:HST}
\end{center}
\end{figure}

For evaluations, a four-\ac{TRP} deployment
at 2 GHz carrier frequency using 20 MHz bandwidth (50 resource blocks) with \ac{TRP} hight of 35m,
30 kHz subcarrier spacing, and train speed \gf{of} 500 km/h is assumed,
as illustrated in Figure \ref{Fig:HST} (upper) is used.
The \ac{TRP} antenna orientation is set to 10 degrees downtilt with an antenna gain of 20.5 dBi.
The channel between each \ac{TRP} and the reception point at the train
is modeled using an extended \ac{CDL} channel model. 
The \ac{SNR} at train position $D1 = 0$ m is 16 dB \gf{in the SFN deployment},
and an \ac{HARQ} scheme with a maximum number of 3 retransmissions is employed,
using a fixed \ac{MCS} with 64-\ac{QAM}, \gf{low-density parity-
check coding} with code-rate = 0.428.

Figure \ref{Fig:HST} (lower) shows the throughput as a function of distance in the different deployment options.
In the baseline \ac{SFN} transmission scheme, the throughput for \ac{UE} locations,
close to midpoint of two \acp{TRP} does not reach the peak throughput of the modulation and coding scheme used.
This is due to the fact that the equivalent channel formed by the combination of \gf{the two dominant} \ac{CDL}
channels with \ac{LoS} components having equal and opposite Doppler shifts results
in a less frequency selective channel with deep fades across some of the \ac{OFDM} symbols in a slot.

This channel behavior can be modified by adding a \ac{TRP}-specific cyclic delay diversity 
which converts the effective channel close to midpoint between \acp{TRP} to a frequency selective channel
without deep fades across \ac{OFDM} symbols.
The throughput improvement with \ac{TRP}-specific \gf{cyclic delay diversity} is shown in the figure.
A precompensation scheme where a \ac{TRP}-specific \ac{DFO} 
compensation is performed also improves the throughput close to the midpoint, as seen in the figure.
However, this scheme requires \ac{TRP}-specific reference signals in order for the train
to estimate the Doppler shifts and additional signaling in the uplink direction to feedback the estimates.

The figure also shows the performance of the \ac{DPS} scheme with genie selection of the \acp{TRP}, 
where a single \ac{TRP} closest to the train is used for data transmission.
In case of \ac{DPS},
the received \ac{SNR} at a train position is smaller than in the case of \ac{SFN} transmissions
due to transmission from a single \ac{TRP}.
However, if sufficient SNR  
can be guaranteed 
using proper deployment,
\ac{DPS} achieves peak throughput of the \ac{MCS} used at all train positions.
\gf{The evaluation results show that a complex scheme, such as \ac{SFN} with \ac{DFO} precompensation,
does not perform significantly better than the other alternatives.
Also, a low-complexity and \ac{UE} transparent scheme such as \ac{SFN} with \ac{TRP}-specific \ac{CDD},
which can be readily employed and scaled to serve a large number of \acp{UE},
performs well by suitably altering the effective channel.}

\subsection{Service-Specific Scheduling}

\begin{figure}[t]
\begin{center}
\includegraphics[width=0.95\columnwidth]{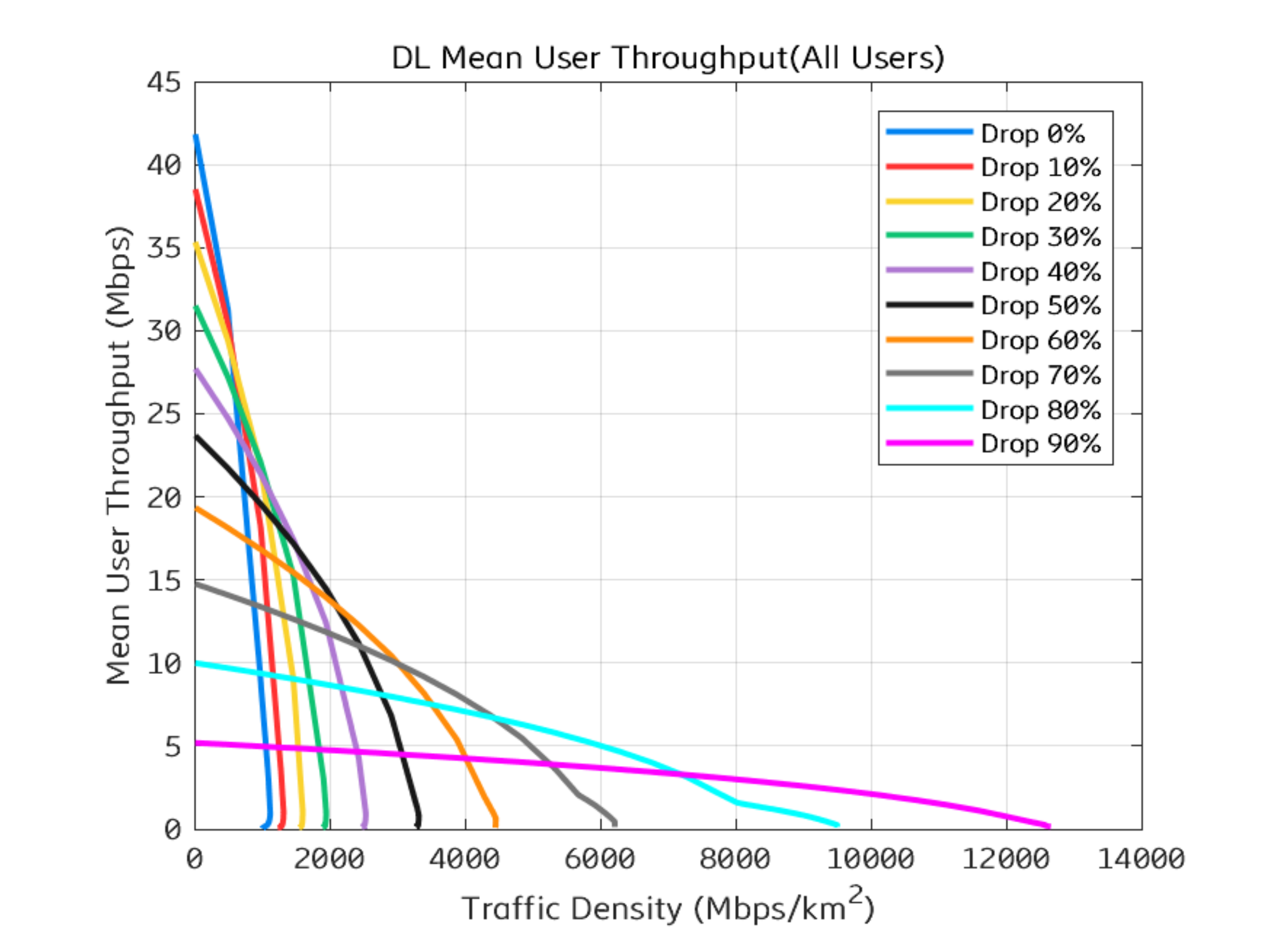}
\caption{Downlink mean user throughput as a function of the traffic density when employing different drop rates
in an automotive highway scenario, in which the base station sites are deployed with 1732m inter-site distance.
}
\label{Fig:Sched}
\end{center}
\end{figure}

To illustrate the impact of customizing the scheduler to operate in specific deployment scenarios,
we consider an automotive high-way scenario, in which the \ac{BS} sites are deployed according to the
3GPP recommendation in \cite{TR36.885}.

Figure \ref{Fig:Sched} shows that dropping low path-gain users improves
the mean user throughput, especially in case of high traffic density.
Specifically, when all users are scheduled simultaneously (marked as 0\% Drop)
the mean user throughput drops to close 0 when the traffic is around or higher than 1000 Mbps/km$^2$.
When the scheduler is configured to distinguish low and high path-gain vehicles and postpones
scheduling vehicles that are momentarily have low path-gain, the average throughput significantly
increases. Dropping 50\% of the low path-gain users (Drop 50\% line), for example,
improves the mean user throughput when traffic density lies between 1000-3000 Mbps/ km2.

This simple example illustrates that adjusting the drop ratio (by adjusting the threshold between
low and high path gain vehicles and setting the drop ratio in the low path-gain group) -- according to
the deployment parameters and prevailing traffic density -- can optimize the system spectral efficiency.
The basic rationale for this is that for non-latency-critical services,
the user data transmission can wait until the user is moving into good coverage area,
while users in poor coverage area can be dropped to improve both system resource utilization
and spectral efficiency.

As an example, consider the highway scenario, in which vehicles drive at 140km/h.
To guarantee 95\% MBB-like service coverage (10 Mbps data rate for \ac{DL} and 2 Mbps for \ac{UL}),
the \ac{DL} capacity for the non-dropping case is 450 Mbps/km$^2$ (760 Mbps/km$^2$ for \ac{UL}).
It takes 50s (250s for UL) to transmit a 500M file in the \ac{DL}.
For the drop 50\% case, for the same traffic density (450 Mbps/km2 for DL and 760 Mbps/km2 for \ac{UL}),
the transmission time is greatly reduced (19s for DL and 31s for \ac{UL}).
Although the Drop 50\% case has about 50\% coverage hole,
this coverage hole lasts until the vehicles drive by a next \ac{BS}, improve their path loss,
and complete the ongoing file transmission.

\subsection{QoS Prediction}

\begin{figure}[t]
\begin{center}
\includegraphics[width=1.0\columnwidth]{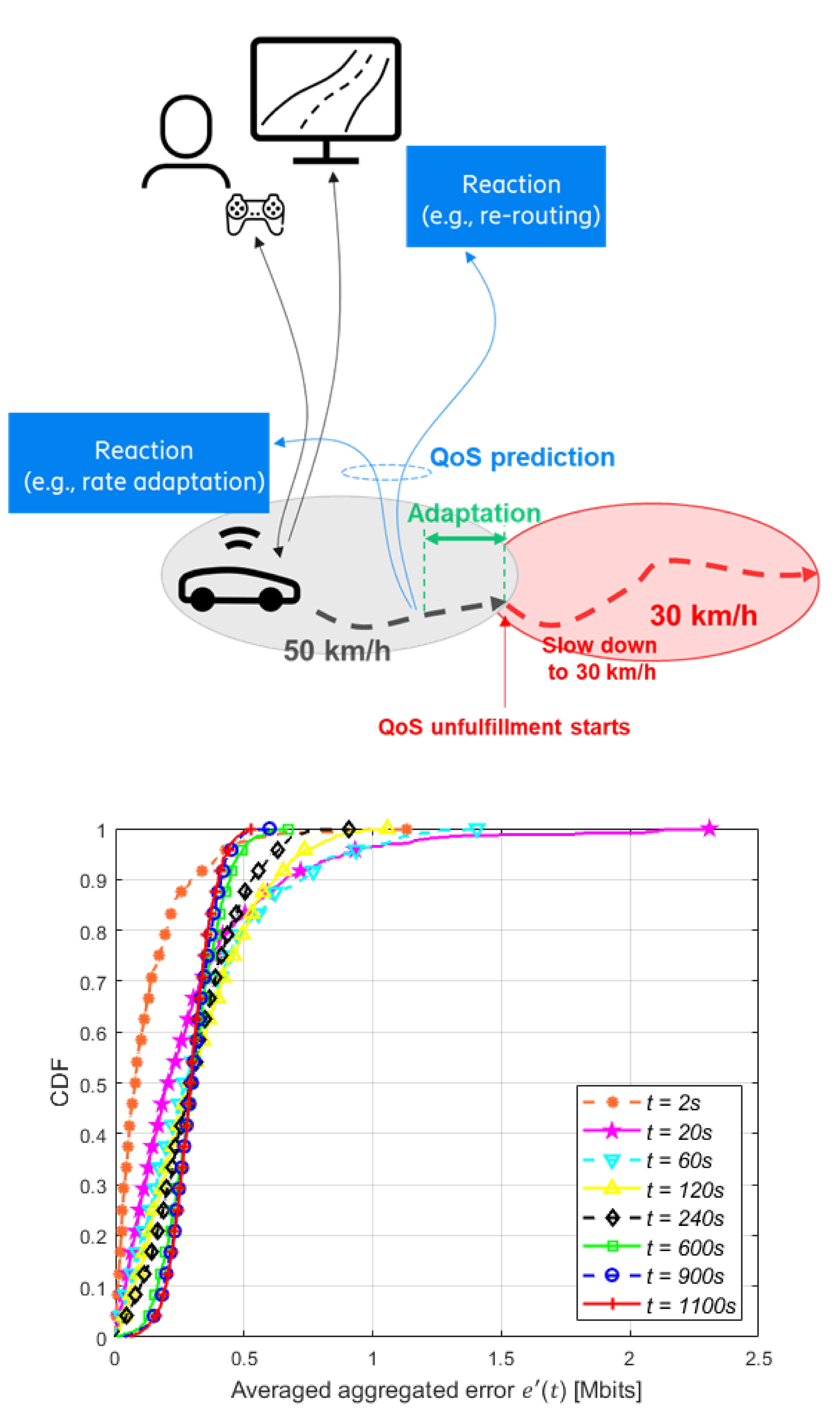}
\caption{Predictive QoS: Linking network-level QoS level and the application layer (upper), Network-based prediction function utilizing \ac{ML} (middle) and \ac{DL} throughput prediction (lower).
}
\label{Fig:pQoS}
\end{center}
\end{figure}

The most fundamental question of predictive \ac{QoS} is, perhaps, the accuracy of the predictions.
We have studied multiple alternatives to predict DL throughput in different time horizons.
Figure~\ref{Fig:pQoS} summarizes our findings in terms of the \ac{CDF} for different prediction horizons
of the difference between predicted and delivered number bits $B$ over a given time interval $\Delta t$:

\begin{align*}
e'(t,\Delta t) & \triangleq \frac{\Big|B_{\text{delivered}}(t)-B_{\text{predicted}}(t)\Big|}{\Delta t}.
\end{align*}

Our results show that both short-term and long-term predictions are quite accurate in most cases.
In the former case, this is due to the ability to predict short term channel quality fluctuations,
while other system variables that are harder to predict (e.g., interference level, instantaneous cell load)
are relatively stable. In the latter case, the longer interval averages out the instantaneous variations
of channel quality and other short-term effects.
In contrast, \ac{DL} throughput prediction in the intermediate
regime is much more challenging. In this case, rapid channel fluctuations are often not well predicted,
while the averaging effect is still weak.
How to bridge the gap between short- and long-term predictions is still an open question.

\section{Concluding Remarks and Outlook}
\label{Sec:Conc}
5G NR was designed to enable various use cases, reach a broad range of aggressive performance targets
and be deployed in both traditional and \ac{mmWave} frequency bands. The initial release (Rel-15) of \ac{NR}
included support for flexible numerology, latency-optimized frame structure, massive \ac{MIMO}, interworking
between low and high frequency bands and dynamic \ac{TDD}. The new features of NR in subsequent releases
include enhancements for MIMO, \ac{V2X}, high-speed \ac{UE}, and \ac{URLLC} services, more accurate positioning,
and support for non-terrestrial and mission-critical communications. These new standardized features,
together with proprietary and algorithmic solutions facilitate connected and intelligent transport services,
including automotive, rail, air transport and public safety services.

Connected and intelligent transport systems will continue to rely on ubiquitous broadband connectivity as
expectations by the automotive, rail and air transport and public safety stakeholders evolve, and new business
models emerge.
The contours of future 6G systems are already emerging, as the standardization and research
communities and end-users in the transport sector define future requirements and solutions.
Based on these
discussions, 6G technology candidates include both further enhancements of existing features and entirely
new features.
The former group includes pushing the limits of frequencies towards the lower bands of THz spectrum,
while examples for the latter include integrated communication and radar sensing, 
integrated terrestrial and non-terrestrial networks, 
and utilizing intelligent reconfigurable surfaces and
full-duplex communications. 
\bibliographystyle{IEEEtran}
\bibliography{D2D5Gv10,IEEEfull}

\end{document}